\documentstyle[eqsecnum,amsfonts,aps]{revtex}
\draft
\begin{document}

\title{CURRENTS, CHARGES, AND CANONICAL STRUCTURE OF  PSEUDODUAL CHIRAL
MODELS }

\author{ Thomas Curtright\thanks{Electronic mail:
curtright@phyvax.ir.Miami.edu, tcurtright@umiami} }
\address{Department of Physics, University of Miami, Box 248046,
 Coral Gables, Florida 33124, USA}

\author{ Cosmas Zachos \thanks{Electronic mail: zachos@hep.anl.gov} }
\address{High Energy Physics Division,
  Argonne National Laboratory,
  Argonne, IL 60439-4815, USA}

\date{January 1994}

\maketitle

\begin{abstract}
We discuss the pseudodual chiral model to illustrate
a class of two-dimensional theories
which have an infinite number of conservation laws
but allow particle production,
at variance with naive expectations.
We describe the symmetries of the pseudodual model,
both local and nonlocal, as transmutations of the symmetries
of the usual chiral model.  We refine the conventional algorithm
to more efficiently produce the
nonlocal symmetries of the model, and we discuss the complete local current
algebra for the pseudodual theory.  We also exhibit the canonical
transformation which connects the usual chiral model to its
fully equivalent dual, further distinguishing the pseudodual theory.
\end{abstract}

\widetext

\section{Introduction}

Many integrable models in two-dimensions have the limiting feature of {\em %
no particle production}. There is a variant of the sigma model for which
this is not so, however, the so-called {\em Pseudodual Chiral Model} of
Zakharov and Mikhailov \cite{Zak}, for which all interactions are distilled
into a very simple, constant torsion term in the lagrangean. The essential
quantum features of the model were first identified by Nappi \cite{nappi},
who calculated the nonvanishing $2\rightarrow 3$ production amplitude for
the model, and who also demonstrated that the model was inequivalent to the
usual Chiral Model in its behavior under the renormalization group: The
Pseudodual Model is not asymptotically free. Although these were lowest
order perturbative calculations carried out for a massless theory, and are
subject to the well-known interpretation problems inherent to a field theory
of massless scalar particles in two-dimensions, it is nonetheless clear that
the physics of the pseudodual model is very different from that of the usual
chiral model, and therefore a full comparison between the two theories is
warranted.

The models were previously compared within the framework of covariant path
integral quantization by Fridling and Jevicki, and similarly by Fradkin and
Tseytlin, \cite{Fridling}. However, the focus of those earlier comparisons
was to exhibit dualized sigma models, with torsion, which were completely
equivalent to the usual sigma model. Indeed, it was shown that a model fully
equivalent but dual to the usual chiral model could be constructed, provided
both nontrivial torsion and metric interactions were included in the
lagrangean.

In this paper, we focus on the differences between the Pseudodual Model and
the usual Chiral Model without enforcing equivalence. We investigate the
pseudodual model at the classical level and within the framework of
canonical quantization, with emphasis on the symmetry structure of the
theory. We consider both local and nonlocal symmetries, and compare with
corresponding structures in the usual chiral model. We exhibit a canonical
transformation which connects the usual chiral model with its fully
equivalent dual version, further clarifying the inequivalence of the
pseudodual theory. We provide a technically refined algorithm for
constructing the nonlocal currents of the pseudodual theory, an algorithm
which is particularly well-suited to models with topological currents for
which the usual recursive algorithm temporarily stalls at the lowest steps
in the recursion before finally producing genuine nonlocals at the third
step and beyond. We also consider in some detail the current algebra for the
full set of local currents in the pseudodual theory, thereby providing an
extension of several recent studies \cite{Forger}. Other, related, more
recent investigations can be found in \cite{balog}.

\section{The Pseudodual Chiral Model}

The familiar two-dimensional chiral model (CM) for matrix-valued fields $g$
is defined by
\begin{equation}
{\cal L}_1=\hbox{Tr}~\partial _\mu g\partial ^\mu g^{-1},
\end{equation}
whose equations of motion are conservation laws
\begin{equation}
\partial _\mu J^\mu =0~~~\Longleftrightarrow ~~~\partial _\mu L^\mu =0.
\end{equation}
Here, $J_\mu \equiv g^{-1}\partial _\mu g$ and$~L_\mu \equiv g\partial _\mu
g^{-1}$, the right- and left-rotation Noether currents of $G_{left}\times
G_{right}$, respectively. E.g., for $G=O(N)$, these currents are
antisymmetric $N\times N$ matrices. We may consider this particular case in
the following discussion without essential loss of generality. Note that $%
{\cal L}_1=-\hbox{Tr}~(J_\mu J^\mu )=-\hbox{Tr}~(L_\mu L^\mu )$. From the
pure-gauge form of the local currents, it follows that the appropriate
non-abelian field strength vanishes:
\begin{equation}
{\partial }_\mu J_\nu -{\partial }_\nu J_\mu +[J_\mu ,J_\nu
]=0~~\Longleftrightarrow ~~\varepsilon ^{\mu \nu }{\partial }_\mu J_\nu
+\varepsilon ^{\mu \nu }J_\mu J_\nu =0~,
\end{equation}
and likewise for $L_\mu $. In this sense, the local currents are
curvature-free. Curvature-free currents generally underlie nonlocal-symmetry
generating algorithms, as will be reviewed in a forthcoming section.

Alternatively, the roles of current conservation and vanishing field
strength may be transmuted. Consider a remarkable transformation \cite
{Zak,nappi} of the ``pseudodual'' type (in the language of \cite{Fridling})
which leads to a drastically different model for an antisymmetric matrix
field $\phi $. For conserved currents, one may always define
\begin{equation}
\label{curl}J_\mu =\varepsilon _{\mu \nu }\partial ^\nu \phi .
\end{equation}
This, of course, is conserved identically. On the other hand, the
curvature-free condition above may now serve instead as the equation of
motion
\begin{equation}
\label{motion}\partial ^\mu \partial _\mu \phi -{\textstyle \frac 12}%
\varepsilon _{\mu \nu }[\partial ^\mu \phi ,\partial ^\nu \phi ]=0,
\end{equation}
which follows from the lagrangean of Zakharov and Mikhailov \cite{Zak}:
\begin{equation}
{\cal L}_2=-{\textstyle \frac 14}\hbox{Tr}~\Bigl(\partial ^\mu \phi \partial
_\mu \phi +{\textstyle \frac 13}\phi \varepsilon _{\mu \nu }[\partial ^\mu
\phi ,\partial ^\nu \phi ]\Bigr).
\end{equation}
This is the definition of the pseudodual chiral model (PCM). Having
transmuted the conservation and curvature conditions obeyed by the local
currents under the interchange CM$\leftrightarrow $PCM, let us also consider
the transmutation of the fundamental symmetries generated by these and other
currents.

Consider first the charges for the local $J_\mu $,
\begin{equation}
Q=\int \!dx\;J_0(x).
\end{equation}
For the currents of the chiral model, the time variation of $Q$ vanishes for
field configurations which extremize ${\cal L}_1$ by Noether's theorem,
while for the currents of the pseudodual model, $Q$ are time-independent for
{\em any} configurations with fixed boundary conditions by merely supposing
the local field $\phi $ is temporally constant at spatial infinity. For the
PCM, $Q=\phi (\infty )-\phi (-\infty )$ is just a topological ``winding'' of
the field onto the spatial line, and thus invariant under the continuous
flow of time \cite{Coleman}.

Noether's procedure does yield results for the pseudodual model, but says
nothing about the previous $Q$. Instead, the $G_{right}$-transformation
invariance of ${\cal L}_2$, $~~\phi \rightarrow O^T\phi O~~$ yields the
(on-shell conserved) Noether currents\footnote{%
Occasionally it may be useful to recall the identities $\varepsilon ^{\kappa
\lambda }\varepsilon ^{\mu \nu }=g^{\kappa \nu }g^{\lambda \mu }-g^{\kappa
\mu }g^{\lambda \nu }{}$ and $g^{\kappa \lambda }\varepsilon ^{\mu \nu
}+g^{\kappa \mu }\varepsilon ^{\nu \lambda }+g^{\kappa \nu }\varepsilon
^{\lambda \mu }=0$.}
\begin{equation}
R_\mu =[\phi ,\tilde J_\mu ]+{\textstyle \frac 13}[\phi ,[J_\mu ,\phi
]]=[\phi ,\partial _\mu \phi ]+{\textstyle \frac 13}\varepsilon _{\mu \nu
}[\phi ,[\partial ^\nu \phi ,\phi ]],
\end{equation}
where $\tilde J_\mu \equiv \varepsilon _{\mu \nu }\,J^\nu $. In contrast to
the chiral model, it is these currents, and not $J_\mu $, which generate
(adjoint) right-rotations in the PCM.

Moreover, as remarked by Nappi \cite{nappi} in footnote (7), the action
specified by the integral of ${\cal L}_2$ is also invariant under the
nonlinear symmetries $\phi\rightarrow \phi +\xi $ whose Noether currents are%
\footnote{%
This transformation preserves the constant-at-infinity boundary conditions
in the PCM, as always in spontaneous symmetry breaking, and, even when the
values of the constant field at $x=\pm \infty $ differ, will not change the
value of the winding $Q$.}
\begin{equation}
Z_\mu =\tilde J_\mu +{\textstyle \frac 12}[J_\mu ,\phi ]=\partial _\mu \phi +%
{\textstyle \frac 12}\varepsilon _{\mu \nu }[\partial ^\nu \phi ,\phi ].
\end{equation}
Indeed, the conservation law for these currents simply amounts to the
equations of motion (\ref{motion}) for the PCM, originally introduced as a
null-curvature condition for the topological $J_\mu $ currents of the model.
Thus the equations of motion have been transmuted from conservation of $%
J_\mu $ for the chiral model to conservation of $Z_\mu $ for the pseudodual
model. These $Z_\mu $ currents are not curvature-free, however, but are
instead J-covariant-curl-free:
\begin{equation}
\label{CovCurlfree}\varepsilon ^{\mu \nu }\partial _\mu Z_\nu +\varepsilon
^{\mu \nu }[J_\mu ,Z_\nu ]=0~.
\end{equation}
In some contrast to ${\cal L}_1$, we note further that ${\cal L}_2=
\frac1{12}\hbox{Tr}~\Bigl(J_\mu (J^\mu +2\tilde Z^\mu )\Bigr)$.

We will demonstrate in Section \S4 below that these ``new'' local conserved
currents, $Z_\mu $ and $R_\mu $, are actually transmutations of the usual
first and second nonlocal currents of the chiral model, respectively. All
three sets of currents, $J_\mu ,Z_\mu ,R_\mu $, transform in the adjoint
representation of $O(N)_{right}$ (the charge of $R_\mu $). By inspection of
the transformation on the field $\phi $, it would appear that the charges
for the shift symmetry {\em commute} among themselves. This is not quite
correct, however, as we shall see in the results of Section \S5 below. In
anticipation of those later results, it turns out that the shift charges
induce a transformation of the $\phi$ conjugate momenta which depends on the
topological charge $Q$: this only vanishes on states with trivial winding.
We shall therefore refer to these shift charges as ``pseudoabelian''.

In ref.\cite{nappi} it was demonstrated how the PCM is radically {\em %
different} from the chiral model, as the former, unlike the latter, is not
asymptotically free, and it allows particle production already at the tree
(semiclassical) level. For sigma models such as the above chiral ones, the
suppression of particle production has been argued (ref.\cite{LuscherZamo})
on the basis of the nonlocal conservation laws of L\"uscher and Pohlmeyer,
ref.\cite{LuscherPohlmeyer}. Nevertheless, we will show that the yet higher
(third and beyond) transmutations of the nonlocal charges of the chiral
model, are, in fact, genuine nonlocal charges for the PCM, and thus the
pseudodual model rather remarkably exhibits {both particle production and an
infinite sequence of nonlocal charges.}

Further note that, properly speaking, the left-invariance $G_{left}$ has
degenerated: for the field $\phi $, left transformations are inert, and thus
right, or axial, or vector transformations are all indistinguishable. The $%
G_{left}\times G_{right}$ symmetry of the chiral model, the axial generators
of which are realized nonlinearly, has thus mutated in the PCM. On the one
hand it has been reduced by the loss of $G_{left}$, but on the other hand it
has been augmented by the nonlinearly realized pseudoabelian $Q_Z$ charges.

The reader may wonder then how the conserved left-currents $L_\mu $ of the
chiral model are realized on the solution set of the PCM. They actually do
not generate left-rotations on the fields $\phi ,$ any more than the $J_\mu $
generate right-rotations. By analogy to the phenomenon detailed in \S4 to
follow, in the PCM the left-currents may be realized nonlocally. This
follows by a direct construction. Upon identifying the right-currents with $%
\varepsilon _{\mu \nu }\partial ^\nu \phi $, one may write $\partial _\mu
g=g~\varepsilon _{\mu \nu }\partial ^\nu \phi $. That is,
\begin{equation}
\partial _1g=g~\partial _0\phi ~,
\end{equation}
which may now be integrated at a fixed time to obtain
\begin{equation}
\label{g-as-fcn-of-phi}g(x,t)=g_0~P\exp (\int_x^\infty \kern%
-1.2emdy~\partial _0\phi (y,t))~,
\end{equation}
assuming the boundary conditions $g(\infty ,t)=g_0$. The left-currents are
now realizable as explicit nonlocal functions of $\phi $ as obtained from
using expression (\ref{g-as-fcn-of-phi}) for $g(x,t)$ to
similarity-transform the right currents:
\begin{equation}
\begin{array}{c}
L_\mu =g\partial _\mu g^{-1}=-g~(g^{-1}\partial _\mu g)~g^{-1}=-g~J_\mu
{}~g^{-1} \\
\\
=-\varepsilon _{\mu \nu }~g~\partial ^\nu \phi ~g^{-1}=-\varepsilon _{\mu
\nu }\partial ^\nu (g~\phi ~g^{-1})+g~[\partial _\mu \phi ~,\phi ]~g^{-1}.
\end{array}
\end{equation}
The first term is trivially conserved while the second one resembles a
similarity-transform of the right currents $R_\mu $, and, in fact, to
leading order, likewise generates adjoint right-rotations sandwiched within
the arbitrary boundary-condition $g_0$ similarity transformation. These
nonlocal currents transform in the adjoint of $G_{left}$, albeit somewhat
speciously, since these transformations only serve to rotate the arbitrary
boundary conditions $g_0$, and do not affect the dynamical fields $\phi $ of
the action at all. They thus commute with the right-rotations. As a
consequence, removing 
$g_0$ from the above currents banishes $G_{left}$ from the theory altogether.

As is evident from the equivalent status of left-versus-right in the chiral
model, none of the above results hinges crucially on the difference between
left- and right-currents. Left$\leftrightarrow $right-reflected identical
results would have followed the above pseudodual transmutation upon
interchange of left with right.

\section{Canonically Equivalent Dual Sigma Model}

The above expression for $g(x,t)$, as an explicitly nonlocal function of$%
\;\partial _0\phi (y,t)$ up to boundary conditions, would seem to suggest
that the two versions of the chiral model are equivalent, providing as it
does an invertible, fixed-time map relating all $g$ and $\phi $ field
configurations. Nevertheless, the point is that this map is{\em \ not }a
canonical transformation. Hence, the $g$ and $\phi $ theories are {\em not}
canonically equivalent. The quantum theories for ${\cal L}_1$ and ${\cal L}%
_2 $ are thus inequivalent, if effects are computed in the standard way, say
in perturbation theory, which assumes canonical variables.

One direct way to see this point would be to compare Poisson brackets for
various expressions in the $g$ and $\phi $ theories. This is done below for
the currents of the $\phi $ theory.

It is instructive, however, to take an indirect approach and construct a
canonical transformation which maps the usual chiral sigma model onto an
equivalent dual sigma model, with torsion, which is different from the PCM.
The transformation identifies conserved, curvature-free currents differently
than in (\ref{curl}). Such a construction is the canonical, fixed-time
analogue of the Lagrange multiplier technique, and accompanying change of
variables, employed by Fridling and Jevicki \cite{Fridling} in the covariant
path integral formalism. We exhibit here such a transformation.

For specificity, consider the standard O(4)~$\simeq $~O(3)$\times $O(3)~$%
\simeq $~SU(2)$\times $SU(2) chiral sigma model, parameterized in the usual
way, $g=\varphi ^0+i\tau ^j\varphi ^j$, with coordinates $\varphi ^0,\varphi
^j~~(j=1,2,3)$, subject to the constraint $(\varphi ^0)^2+{\bf \varphi }^2=1$%
, with ${\bf \varphi }^2\equiv $ $\sum_{j\;}(\varphi ^j)^2$. Resolve the
constraint and substitute $\varphi ^0=\pm \sqrt{1-{\bf \varphi }^2}$, to
obtain the standard form for the defining chiral lagrangean
\begin{equation}
{\cal L}_1={\textstyle\frac12}\left( \delta ^{ij}+\frac{\varphi^i\varphi^j}
{1-{\bf \varphi }^2}\right) \partial _\mu \varphi ^i\partial ^\mu \varphi ^j.
\end{equation}
We now show that this model is {\em canonically equivalent} to the dual
sigma model (DSM) defined by the lagrangean\footnote{%
Up to normalizations, this is essentially the lagrangean of
Fridling and Jevicki, as in ref.\cite{Fridling},
although the reader will note some significant sign differences
from their Eqn(13).}
\begin{equation}
{\cal L}_3=\frac 1{1+4{\bf \psi }^2}\left( {\textstyle\frac12}
\left( \delta ^{ij}+4\psi
^i\psi ^j\right) \partial _\mu \psi ^i\partial ^\mu \psi ^j-\varepsilon
^{\mu \nu }\varepsilon ^{ijk}\psi ^i\partial _\mu \psi ^j\partial _\nu \psi
^k\right) .
\end{equation}
Of course, this lagrangean is different than that for the PCM, ${\cal L}_2$,
since it contains both nontrivial metric and torsion on the field manifold.
Nonetheless, the reader is invited to compare the two to leading orders in
the fields $\psi $ and $\phi $, respectively, i.e.~to examine the weak $\psi
$ field limit in the above and in what follows.

As we shall verify in detail, a suitable generator for a canonical
transformation relating $\varphi $ and $\psi $ is simply given by a
spacelike line integral of an O(3) invariant bilinear $\psi ^iJ_i^\mu
[\varphi ,\varpi ]$, where $J_i^\mu $ is either the left or the right,
conserved, curvature-free O(3) current for the $\varphi $ theory. Choosing
here the right current ($V+A$), at any fixed time our generator is $F[\psi
,\varphi ]=\int_{-\infty }^\infty dx\;\psi ^iJ_i^1[\varphi ]$, as given
explicitly by
\begin{equation}
\label{F}F[\psi ,\varphi ]=\int_{-\infty }^{+\infty }dx~\psi ^i\;\left(
\sqrt{1-{\bf \varphi }^2}\frac \partial {\partial x}\varphi ^i-\varphi
^i\frac \partial {\partial x}\sqrt{1-{\bf \varphi }^2}+\varepsilon
^{ijk}\varphi ^j\frac \partial {\partial x}\varphi ^k\right) .
\end{equation}
Note that left-rotations on $\varphi $ alone do nothing to this $F$. It is
then consistent to assume that the new field variable $\psi ^i$ is a
left-transformation singlet, just like its conjugate quantity $J_i^1[\varphi
]$, and that $F[\psi ,\varphi ]$ is left-invariant. On the other hand, the
effects of right-rotations (or separate axial or vector rotations) on $%
\varphi $ must be compensated by appropriate isospin transformations of $%
\psi $ analogous to the ones generated by $R_\mu $ for the PCM in the
previous section.\footnote{%
Although, as discussed below, the actual curvature-free current for $\psi $
is a combination of isospin and topological currents, so when comparing the
effect of charges for the curvature-free currents between $\varphi $ and $%
\psi $ theories, there is a possibility of subtle surface term contributions
reminiscent of the constants of integration $g_{0}$ that appeared in (\ref
{g-as-fcn-of-phi}).}

By construction, $F$ is linear in $\psi $, but nonlinear in $\varphi $. Note
that integrating by parts just gives back the original $F$, without any
surface term. Also, while we have written $\pm \infty $ as the limits of $x$
integration in the expression for $F$, the reader should be aware that
finite limits of integration are also acceptable with appropriate boundary
conditions. For example, $x$ could be a circle, with both $\psi $ and $%
\varphi $ satisfying periodic boundary conditions. Finally, note that the
weak $\varphi $ field limit of the generating function reduces to the
well-known duality transformation between free scalar and pseudoscalar
fields, as generated by $F_0[\psi ,\varphi ]=\int dx~\psi ^i\frac \partial
{\partial x}\varphi ^i$.

Having exhibited the canonical transformation which relates the chiral model
to its fully equivalent dual, we may now allow modifications in the form of $%
F$ to see if it is also possible to connect the chiral model to the PCM. The
results of such an investigation are negative: The nontrivial metric of the
chiral model cannot be converted into the trivial metric and constant
torsion of the PCM through canonical transformations which equate the
curvature-free currents of the two theories. This follows from explicit
calculations similar to those below. It should not be difficult, in
principle, for a sufficiently motivated reader to see this, say by allowing
arbitrary invariant functions to appear in $F$. Nevertheless, the details
can be tedious, and shortcuts are not available, at present. We forego the
details here and simply state the result.

Also, in contrast to the emergent nonlinear pseudoabelian Z-charges in the
PCM model, no such local symmetry appears present in the DSM. Thus, there is
no {\em local} analog of the three axial $\varphi $ symmetries for the
theory ${\cal L}_3$. It is straightforward to verify this for the classical
theory, and it is best left as an exercise. (At the quantum level, these
facts are consistent with the properties of the transformation functional
discussed below.)

Let us now verify that $F$ generates a canonical transformation which
identifies the curvature-free currents of the two models defined by ${\cal L}%
_1$ and ${\cal L}_3$. As a consequence, it follows that the energy-momentum
tensors for the two theories are equal under the transformation. It is a
textbook exercise to check that $F$ fulfills its mission at the classical
level. Classically, functionally differentiating $F$ generates the conjugate
momenta. The conjugate of $\psi ^i$ is given by%
\begin{eqnarray}
\label{Pi} \pi _i=\frac{\delta F[\psi ,\varphi ]}{\delta \psi ^i}&=&
\sqrt{1-{\bf \varphi }^2}\;\frac \partial {\partial x}\varphi ^i
-\varphi ^i\frac \partial
{\partial x}\left( \sqrt{1-{\bf \varphi }^2}\right) +\varepsilon
^{ijk}\varphi ^j\frac \partial {\partial x}\varphi ^k \nonumber \\
&=&\left( \sqrt{1-{\bf \varphi }^2}\;\delta ^{ij}+\frac{\varphi
^i\varphi ^j}{\sqrt{1-{\bf \varphi }^2}}-\varepsilon ^{ijk}\varphi ^k\right)
\frac \partial {\partial x}\varphi ^j.
\end{eqnarray}
The generator $F$\ was, in fact, chosen to yield this result. The conjugate
of $\varphi ^i$ is obtained through
\begin{eqnarray}
\label{PrelimVarPi}
\varpi _i=-\frac{\delta F[\psi ,\varphi ]}{\delta \varphi ^i}
&=&\left( \sqrt{1-
{\bf \varphi }^2}\;\delta ^{ij}+\frac{\varphi ^i\varphi ^j}{\sqrt{1-{\bf %
\varphi }^2}}+\varepsilon ^{ijk}\varphi ^k\right) \frac \partial {\partial
x}\psi ^j \nonumber \\
&+&\left( \frac 2{\sqrt{1-{\bf \varphi }^2}}\left( \varphi
^i\psi ^j-\psi ^i\varphi ^j\right) -2\varepsilon ^{ijk}\psi ^k\right) \frac
\partial {\partial x}\varphi ^j,
\end{eqnarray}
a ``mixed'' expression involving both $\varphi $ and $\psi $. Now (\ref{Pi})
may be inverted to replace $\frac \partial {\partial x}\varphi $ in (\ref
{PrelimVarPi}) by $\pi $. Thus%
\begin{eqnarray}
\label{VarPi}
\varpi _i&=&\left( \sqrt{1-{\bf \varphi }^2}\;\delta ^{ij}+\frac{%
\varphi ^i\varphi ^j}{\sqrt{1-{\bf \varphi }^2}}+\varepsilon ^{ijk}\varphi
^k\right) \frac \partial {\partial x}\psi ^j \nonumber \\
&+&2\left( \psi ^k\varphi ^k\delta ^{il}-\psi ^i\varphi ^l\right)
\pi _l+2\left( \sqrt{1-{\bf \varphi }^2}\;\delta ^{ij}+\frac{\varphi
^i\varphi ^j}{\sqrt{1-{\bf \varphi }^2}}\right) \varepsilon ^{jkl}\psi ^k\pi
_l ~.
\end{eqnarray}
If viewed as classical relations, we may substitute for $\pi _i$ and $\varpi
_i$, in (\ref{Pi}) and (\ref{VarPi}), in terms of $\frac \partial {\partial
t}\varphi ^j$ and $\frac \partial {\partial t}\psi ^j$, as follows from the
lagrangeans ${\cal L}_1$ and ${\cal L}_3$:
\begin{equation}
\pi _i=\frac 1{1+4{\bf \psi }^2}\left( \left( \delta ^{ij}+4\psi ^i\psi
^j\right) \frac \partial {\partial t}\psi ^j+2\varepsilon ^{ijk}\psi ^j\frac
\partial {\partial x}\psi ^k\right) ,\;\;\;\;\;\varpi _i=\left( \delta ^{ij}+%
\frac{\varphi ^i\varphi ^j}{1-{\bf \varphi }^2}\right) \frac \partial
{\partial t}\varphi ^j.
\end{equation}
The resulting covariant pair of first-order, nonlinear, partial differential
equations for $\varphi $ and $\psi $ constitute a B\"acklund transformation
connecting the two theories. Consistency of this B\"acklund transformation
is equivalent to the classical equations of motion for $\varphi $ and $\psi $%
{}.

Note, as a consequence of (\ref{Pi}), we find the mixed inner product
relation
\begin{equation}
\varphi ^i\pi _i=-\frac \partial {\partial x}\left( \sqrt{1-{\bf \varphi }^2}%
\right) .
\end{equation}
Integrating over all $x$, we obtain a mixed ``adiabatic'' (topological)
invariant, which represents a nontrivial global constraint obeyed by the
correlated pair of solutions for the two theories that are connected by the
canonical map. For example, assuming trivial or periodic boundary conditions
on $\varphi $ , we have $\int dx~\varphi ^i\pi _i=0$.

Now, in the DSM, what is the conserved, curvature-free current? In contrast
to the PCM, where it was essentially {\em forced} to be a topological
current, here a topological current by itself will not suffice. Neither will
a conserved Noether current. For instance, under the isospin transformation $%
\delta \psi ^i=\varepsilon ^{ijk}\psi ^j\omega ^k$, the conserved Noether
current of ${\cal L}_3$ is defined as usual by $I_i^\mu =\delta {\cal L}%
_3/\delta (\partial _\mu \omega ^i)$ so that $I_i^0=\varepsilon ^{ijk}\psi
^j\pi _k$. But this is not curvature free. Instead, the conserved,
curvature-free current ${\cal J}_i^\mu $, to be compared with $J_i^\mu $ of
the CM, is a mixture of this Noether current and a topological current: $%
{\cal J}_i^\mu =2I_i^\mu -\varepsilon ^{\mu \nu }\partial _\nu \psi ^i$ so
that ${\cal J}_i^1=\pi _i$. In covariant form,
\begin{equation}
\label{PsiCurrent}{\cal J}_i^\mu =\frac{-1}{1+4{\bf \psi }^2}\left( \left(
\delta ^{ij}+4\psi ^i\psi ^j\right) \varepsilon ^{\mu \nu }\partial _\nu
\psi ^j+2\varepsilon ^{ijk}\psi ^j\partial ^\mu \psi ^k\right) .
\end{equation}
So, to complete verification that the canonical $F$-generated transformation
does the job, at least classically, it remains to show that the
curvature-free currents\footnote{%
Our normalizations here are such that ``curvature-free current'' means $%
\varepsilon _{\mu \nu }\left( \partial ^\mu J_i^\nu +\varepsilon
^{ijk}J_j^\mu J_k^\nu \right) =0$.} are equal for the $\varphi $ and $\psi $
theories when expressed in terms of the respective fields and their
conjugate variables. That is, we must show $J_i^\mu [\varphi ,\varpi ]={\cal %
J}_i^\mu [\psi ,\pi ].$ This requires us to establish the following:
\begin{equation}
\label{Jspace}{\cal J}_i^1\equiv \pi _i=\left( \sqrt{1-{\bf \varphi }^2}%
\;\delta ^{ij}+\frac{\varphi ^i\varphi ^j}{\sqrt{1-{\bf \varphi }^2}}%
-\varepsilon ^{ijk}\varphi ^k\right) \frac \partial {\partial x}\varphi
^j\equiv J_i^1,
\end{equation}
\begin{equation}
\label{Jtime}{\cal J}_i^0\equiv -\frac \partial {\partial x}\psi
^i-2\varepsilon ^{ijk}\psi ^j\pi _k=-\sqrt{1-{\bf \varphi }^2}\;\varpi
_i-\varepsilon ^{ijk}\varphi ^j\varpi _k\equiv J_i^0.
\end{equation}
The first of these is precisely (\ref{Pi}), the form for $\pi_i$ given
directly by the generating functional.

The second (time component) current identity takes more effort to establish,
but it also follows from (\ref{Pi}), and (\ref{VarPi}). To see this, reduce
each of the expressions involving the fields and their conjugate momenta in (%
\ref{Jtime}) to the following mixed result involving the fields and their
spatial derivatives.
\begin{equation}
{\cal J}_i^0=-\frac \partial {\partial x}\psi ^i+2\psi ^j\left( \varphi
^j\frac \partial {\partial x}\varphi ^i-\varphi ^i\frac \partial {\partial
x}\varphi ^j\right) +2\varepsilon ^{ijk}\psi ^j\left( \varphi ^k\frac
\partial {\partial x}\left( \sqrt{1-{\bf \varphi }^2}\right) -\sqrt{1-{\bf %
\varphi }^2}\frac \partial {\partial x}\varphi ^k\right) =J_i^0.
\end{equation}
Thus, the validity of our canonical transformation at the classical level is
now established. What about the ensuing relation between the quantum
theories associated with ${\cal L}_1$ and ${\cal L}_3$?

To carry out a comparison at the quantum level, we combine Schr\"odinger
wave-functional methods with the transformation theory of Dirac, as
explained, for example, within the context of Liouville theory in \cite
{CurtrightGhandour}. We find that the energy-momentum eigenfunctionals of
the quantum $\psi $ theory, $\Psi _{E,\vec p}[\psi ]$, are related to those
of the quantum $\varphi $ theory, $\Phi _{E,\vec p}[\varphi ]$, and vice
versa, by exponentiating the classical generating function of the canonical
transformation to obtain a transformation functional. That is, at any fixed
time\footnote{%
Since this is a fixed-time expression, the Schr\"odinger functional integral
$\int d\varphi \;$is over all field configurations at each point in space,
but not at any other times. That is, $\int d\varphi $ in (\ref
{eigenfunctionals}) is {\em not} a path integral.} the eigenfunctionals of
the two theories are nonlinear (in $\varphi $) functional Fourier transforms
of one another.
\begin{equation}
\label{eigenfunctionals}\Psi _{E,\vec p}[\psi ]=N\;\int d\varphi
\;e^{iF[\psi ,\varphi ]}\;\Phi _{E,\vec p}[\varphi ].
\end{equation}
Note that we have allowed for an adjustment of the overall normalization of
eigenfunctionals in the two theories by including an undetermined
energy--momentum dependent factor, $N$. In principle, it is straightforward
to determine $N$, but we have not.

The previous classical relations between the fields and their conjugate
momenta, or alternatively, between the curvature-free currents, now become
nonlinear first-order functional differential equations obeyed by the
transformation functional:
\begin{eqnarray}
{\cal J}_i^1[\pi \equiv  -i\frac \delta {\delta \psi }]\;e^{iF[\psi ,\varphi
]}&=&J_i^1[\varphi ]\;e^{iF[\psi ,\varphi ]},
\\
{\cal J}_i^0[\psi ,\pi \equiv  -i\frac \delta {\delta \psi }]\;e^{iF[\psi
,\varphi ]}&=&-J_i^0[\varphi ,\varpi \equiv -i\frac \delta {\delta \varphi
}]\;e^{iF[\psi ,\varphi ]}~. \nonumber
\end{eqnarray}
Presumably these differential equations are exact, and the transformation
functional provides an exact solution of them. Imposing a cut-off on the
theory, it is straightforward to check that this is indeed so simply as a
consequence of the classical equations (\ref{Jspace}) and (\ref{Jtime}). (In
this regard, we note that dimensional regularization is particularly
convenient here, as well-known for the CM, since it permits us to naively
discard certain $\delta (0)$ terms which would otherwise mar the hermiticity
of $F$ and the currents as written, and which would also confront one in
comparing the energy-momentum tensors for the $\psi$ and $\varphi$
theories.) However, a nontrivial analysis is required to remove the cut-off,
that is, to renormalize the transformation in (\ref{eigenfunctionals}). This
renormalization analysis will not be given here.

Rather, here we take the expression (\ref{eigenfunctionals}) as is and use
it to gain a better understanding of how symmetries in one {\em quantum}
theory are related to those of the other {\em quantum} theory, a technique
previously illustrated in the context of simple potential models \cite
{CurtrightTahoe}. From the transformation properties of the $\phi$-dependent
current in $F$, we expect that $V$-and-$A$-transformations on the
eigenfunctionals must coincide insofar as $e^{iF[\psi ,\varphi ]}$ projects
onto the left-invariants. This behavior for the quantized canonically
equivalent dual sigma model would correspond, in the functional framework,
to the degeneration of $G_{left}$ for the classical PCM illustrated in the
previous section. Indeed, when acting on the above transformation
functional, and consequently also when acting on the eigenfunctionals, the
axial and vector charges of the CM ($\varphi $ theory) produce equivalent
effects on the $\psi$ theory. Both the inhomogeneous, nonlinearly realized
axial symmetries and the linear vector O(3) symmetry of the chiral $\varphi $
theory are projected by the canonical transformation into the same linear
right-isospin O(3) symmetry of the ${\cal L}_3$ theory.
\begin{eqnarray}
\left( \int_{-\infty }^\infty dy\;\sqrt{1-{\bf \varphi }^2(y)}\frac \delta
{\delta \varphi ^i(y)}\right) e^{iF[\psi ,\varphi ]}&=&
-\;\left( \int_{-\infty
}^\infty dy\;\varepsilon ^{ijk}\varphi ^j(y)\frac \delta {\delta \varphi
^k(y)}\right) e^{iF[\psi ,\varphi ]}= \nonumber \\
&=&-\;\left( \int_{-\infty }^\infty dy\;\varepsilon ^{ijk}\psi ^j(y)\frac
\delta {\delta \psi ^k(y)}\right) e^{iF[\psi ,\varphi ]}.
\end{eqnarray}
To take the last step, one must integrate by parts and either cancel or
discard surface contributions. The action of the charges on the
eigenfunctionals, as related by (\ref{eigenfunctionals}) and its inverse, is
obtained by functionally integrating by parts after acting on $e^{iF[\psi
,\varphi ]}$. (Once again, dimensional regularization is convenient here as
it allows us to blithely discard $\delta (0)$ terms.)

The only vestige of the difference in vector and axial symmetries carried by
the transformation functional lies in its field parity properties. Under $%
\psi \rightarrow -\psi $, the generator $F$ trivially changes sign. But
under $\varphi \rightarrow -\varphi $, not only is the sign of $F$ changed,
but also the right current for the CM ($\varphi $ theory) is converted into
the left current, and thus $F[-\psi,-\varphi]$ generates a canonical
transformation which projects onto right-invariants. Therefore, it is
possible to obtain all the effects of interchanging right $\leftrightarrow $
left currents in the transformation functional by merely splitting all wave
functionals into components even and odd under field parity\footnote{%
The same statement applies to interchanging northern and southern
hemispheres for the three-sphere defined by $(\varphi ^0)^2+{\bf \varphi }%
^2=1 $, since this only flips the sign of the square-root appearing in $F$
which is again tantamount to $\varphi \rightarrow -\varphi .$}. Since this
can always be done, using one current in $F$ instead of the other results in
no loss of information.

We now return to our investigation of the PCM itself. In particular, we
systematically study the nonlocal currents and their charges which are
guaranteed to exist for the PCM by virtue of the conservation and vanishing
curvature for $J_\mu $.

\section{Nonlocal Currents and Charges for the Pseudodual Model}

The full set of nonlocal conservation laws is neatly described using the
methods in \cite{CurtrightZachos'80} (see also \cite
{brezin,Polyakov'80,Zachos'80,CurtrightZachos'93}). For any conserved,
curvature-free currents such as $J_\mu $, irrespective of the specific model
considered, introduce Pohlmeyer's dual boost spectral parameter $\kappa $ to
define
\begin{equation}
\label{C}C_\mu (x,\kappa )=-\frac{\kappa ^2}{1-\kappa ^2}J_\mu -\frac \kappa
{1-\kappa ^2}\;\tilde J_\mu ,
\end{equation}
where $\tilde J_\mu \equiv \varepsilon _{\mu \nu }\,J^\nu $ . Thus
\begin{equation}
\label{ZeroCurvC}\;\left( \partial ^\mu +C^\mu \right) \widetilde{C}_\mu
=0\;.
\end{equation}
This serves as the consistency condition for the equations
\begin{equation}
\label{PathOrdExp}\partial _\mu \,\chi ^{ab}(x)=-C_\mu ^{ac}\,\chi
^{cb}(x)\;,
\end{equation}
or, equivalently,
\begin{equation}
\varepsilon _{\mu \nu }\partial ^\nu \chi =\kappa ~(\partial _\mu +J_\mu
)~\chi \;,
\end{equation}
which are solvable recursively in $\kappa $ \cite{brezin,CurtrightZachos'80}%
. Equivalently, the solution $\chi $ can be expressed as a path-ordered
exponential (Polyakov's path-independent disorder variable) \cite
{Polyakov'80,CurtrightZachos'80}
\begin{equation}
\label{realPathOrdExp}\chi (x,\kappa )=P\exp \Bigl(-\int_{-\infty }^x\kern%
-1.2emdy~C_1(y,t)\Bigr)\equiv 1\kern-0.36em\llap~1+\sum_{n=0}^\infty \kappa
^{n+1}\chi ^{(n)}\;.
\end{equation}

These ensure conservation of an {\em antisymmetrized} nonlocal ``master
current'' constructed as follows,
\begin{equation}
\label{NMC}{\frak J}^\mu (x,\kappa )\equiv \frac 1{2\kappa
}~\varepsilon ^{\mu \nu }\partial _\nu \;\Bigl(\chi (x,\kappa )-\chi
^T(x,\kappa )\Bigr)\equiv \sum_{n=0}^\infty \kappa ^n~J_{(n)}^\mu (x)\;.
\end{equation}
The conserved master current acts as the generating functional of all
currents $J_{(n)}^\mu $ (separately) conserved order-by-order in $\kappa $.
E.g.~the lowest four orders yield:
\begin{eqnarray}
{\frak J}_\mu(x,\kappa) &= & J_\mu(x)~
+~\kappa \Biggl( \tilde J_\mu(x)+{\textstyle \frac 12}
\Bigl[J_\mu(x)~,~\int_{-\infty}^x \kern-1.2em dy~ J_0(y)\Bigr] \Biggr)~+
\nonumber  \\
& &+~\kappa^2 \Biggl( \tilde J^{(1)}_\mu(x)+
{\textstyle \frac 12} (J_\mu(x)~\chi^{(1)} +
\chi^{(1)T}~J_\mu(x) ) \Biggr)~+ \nonumber  \\
& &+~\kappa^3 \Biggl( \tilde J^{(2)}_\mu(x)+
{\textstyle \frac 12}(J_\mu(x)~\chi^{(2)} +
\chi^{(2)T}~J_\mu(x)) \Biggr)~+~{\cal O}(\kappa^4)~.
\end{eqnarray}
Integrating the nonlocal master current yields a conserved ``master charge''
\begin{equation}
\label{WeDon'tTakeAmericanExpress}{\frak G}(\kappa )=\int_{-\infty
}^{+\infty }\kern-1.2emdx~{\frak J}_0(x,\kappa )\equiv
\sum_{n=0}^\infty \kappa ^n~Q_{(n)}\;.
\end{equation}
$Q_{(0)}$ is the conventional symmetry charge, while $%
Q_{(1)},Q_{(2)},Q_{(3)},...$ are the well-known nonlocal charges, best
studied for sigma models \cite
{LuscherPohlmeyer,LuscherZamo,brezin,Polyakov'80}, the Gross-Neveu model
\cite{CurtrightZachos'81}, and supersymmetric combinations of the two \cite
{CurtrightZachos'80}.

For the PCM \cite{Zak,nappi}, however, it readily follows from
\begin{equation}
J_\mu ^{(0)}=J_\mu =\varepsilon _{\mu \nu }\partial ^\nu \phi
,~~~\Longrightarrow ~~~\chi ^{(0)}(x)=\phi (x)-\phi (-\infty ),
\end{equation}
that
\begin{equation}
J_\mu ^{(1)}=\partial _\mu \phi +{\textstyle \frac 12}\varepsilon _{\mu \nu
}[\partial ^\nu \phi ,\phi ]-{\textstyle \frac 12}[J_\mu ,\phi (-\infty
)]=Z_\mu -{\textstyle \frac 12}[J_\mu ,\phi (-\infty )].
\end{equation}
Recall that, as stressed in Section \S 2, $\phi (-\infty )$ is taken to be
time-independent, and thus each piece of this current is separately
conserved. So, the CM$\leftrightarrow $PCM transmutation has yielded a local
current for the first ``nonlocal''. Continuing,
\begin{equation}
\chi ^{(1)}(x)=\int_{-\infty }^x\kern-1.2emdy~\Bigl(\partial _0\phi
(y)+\partial _1\phi (y)~\phi (y)\Bigr)-\phi (x)\phi (-\infty )+\phi (-\infty
)^2~.
\end{equation}
Likewise,
\begin{eqnarray}
\!\! J_\mu^{(2)}&=&
\varepsilon_{\mu\nu} \partial^\nu \phi +[\partial_\mu \phi,\phi]
-\phi~ \varepsilon_{\mu\nu}\partial^\nu\phi~  \phi +
{\textstyle \frac 12}\varepsilon_{\mu\nu}\partial^\nu
\Bigl(\phi~ \chi^{(1)} + \chi^{(1)T} \phi \Bigr)\nonumber\\   & &
{}~~-{\textstyle \frac 12}[Z_\mu, \phi(-\infty)] +
{\textstyle \frac 14}\varepsilon_{\mu\nu}\partial^\nu (
\phi^2 \phi(-\infty)+\phi(-\infty)\phi^2 ) ~~~~~~~~~~= \\
&=&
J_\mu -R_\mu -{\textstyle \frac 13} \varepsilon_{\mu\nu}\partial^\nu(\phi^3)~+
{\textstyle \frac 12}\varepsilon_{\mu\nu}\partial^\nu
\Bigl(\phi\chi^{(1)} + \chi^{(1)T} \phi \Bigr)
-{\textstyle \frac 12}[Z_\mu, \phi(-\infty)] +
{\textstyle \frac 14}\varepsilon_{\mu\nu}\partial^\nu (
\phi^2 \phi(-\infty)+\phi(-\infty)\phi^2 ) .\nonumber
\end{eqnarray}
On-shell properties of the currents have been used. However, this second
``nonlocal'' current is also effectively local: the skew-gradient term,
which might appear to contribute a nonlocal piece to the charge via $\chi
^{(1)}$, only contributes $[\phi (\infty ),Q_Z]/2$, i.e.~a trivial piece
based on a local current.

In contrast to the first two steps, however, the third step in the recursive
algorithm gives
\begin{equation}
J_\mu ^{(3)}={\textstyle \frac 12}\Bigl(Z_\mu \chi ^{(1)}+\chi ^{(1)T}Z_\mu
\Bigr)+...~,
\end{equation}
where ellipses (...) indicate terms which contribute only local pieces to
the corresponding charge, whereas the term written explicitly may be seen to
contribute ineluctable nonlocal pieces to that charge. Thus $J_\mu ^{(3)}$
appears, like all higher currents, genuinely non-local. In fact, we will see
below that the action of $Q^{(3)}$ on the field changes the boundary
condition at $x=\infty $ to a different one than at $-\infty $, and thereby
switches its topological sector, which is quantified by $Q^{(0)}$.

{In summary for the pseudodual model, the charge $Q^{(0)}$ is topological,
while $Q^{(1)}$ generates shifts, $Q^{(2)}$ generates `right' rotations, and
$Q^{(n\geq 3)}$ appear genuinely nonlocal.}

It may well be objected that the above master current construction is best
suited to the chiral model, but is not really handy for the PCM, since the
procedure starts off with a non-Noether (topological) current, then
``stalls'' twice at the first two steps before finally producing genuine
non-locals at the third step and beyond. The construction also produces a
clutter of total derivatives and other terms whose charges are extraneous.
To overcome some of these objections, we offer here a refined algorithm
which begins with the lowest nontopological (Noether) current $Z_\mu $ to
produce an alternate but equally viable conserved master current. This
refined construction helps to reduce some of the clutter and only stalls
once. Following \cite{CurtrightZachos'93} as illustrated above, define
\begin{equation}
W_\mu (x,\kappa )\equiv Z_\mu +\kappa \tilde Z_\mu ,
\end{equation}
which is readily seen to be C-covariantly conserved:
\begin{equation}
\partial ^\mu W_\mu +[C^\mu ,W_\mu ]=0~.
\end{equation}
This condition then empowers $W_\mu $ to serve as the seed for a new and
improved conserved master-current
$$
{\frak W}_\mu (x,\kappa )=\chi ^{-1}W_\mu ~\chi =
$$
\begin{equation}
=Z_\mu +\kappa \left(
T_\mu -[Z_\mu ,\phi (-\infty )]\right) +\kappa ^2\left( N_\mu -[T_\mu ,\phi
(-\infty )]+\frac 12[[Z_\mu ,\phi (-\infty )],\phi (-\infty )]\right) +{\cal %
O}(\kappa ^3),
\end{equation}
where we have introduced the convenient combination
\begin{equation}
\label{Tcurrent}T_\mu \equiv J_\mu -{\textstyle \frac 32}R_\mu =\tilde Z_\mu
+[Z_\mu ,\phi ],
\end{equation}
and where now the terms of second order and higher are genuinely nonlocal;
e.g.
\begin{equation}
\label{Ncurrent}N_\mu =[T_\mu ,\phi ]-{\textstyle \frac 12}[[Z_\mu
,\phi ],\phi ]+[Z_\mu ,\int_{-\infty }^x\!\!dy~Z_0(y)]\;.
\end{equation}
This is a refined equivalent of $J_\mu ^{(3)}$ above. Note the terms in $%
{\frak W}_\mu $ involving the constant matrices $\phi (-\infty )$ are
separately conserved, as remarked previously.

In general, it is straightforward to see that the seeds for such improved
master currents only need be conserved currents, such as $Z_\mu $ above,
which also have a vanishing J-covariant-curl of the type (\ref{CovCurlfree}%
). For instance, the previous nonlocal currents themselves may easily be
fashioned to satisfy (\ref{CovCurlfree}), and thereby seed respective
conserved master currents.

As an aside, in some models, such as the CM, master currents such as $%
{\frak W}$ amount to the currents of nonlocal-similarity-transformed
fields, which obey the classical equations of motion if the fields do \cite
{Zachos'80}, thereby allowing an interpretation of the similarity transform
as an auto-B\"acklund transformation (quite distinct from the B\"acklund
transformation of the previous section). Here, however, the corresponding
transform of $\phi $
\begin{equation}
\label{NonlocalPhi}\Phi \equiv \chi ^{-1}~\phi ~\chi ,\;
\end{equation}
does not quite obey the original classical equations of motion. Rather,
\begin{equation}
\label{bigmotion}\partial ^\mu \partial _\mu \Phi -\varepsilon ^{\mu \nu
}\partial _\mu \Phi \partial _\nu \Phi =({\textstyle {\frac{2\kappa }{%
1-\kappa ^2}}})^2~{{\frak W}}^\mu \tilde {{\frak W}}_\mu ,
\end{equation}
which would seem to obviate any interpretation of (\ref{NonlocalPhi}) as an
auto-B\"acklund transformation for the PCM. Nonetheless, (\ref{bigmotion})
is still an evocative relation which encodes all solutions of the PCM.

As an additional aside, it may be worth recalling with ref.\cite{Zak} that
there is also a {\em local }sequence of conserved currents predicated on
conserved, curvature-free currents such as $J_\mu $. This follows directly
from rewriting
\begin{equation}
(\partial _0\pm \partial _1)(J_0\mp J_1)=\pm {\textstyle \frac 12}%
[J_0-J_1,J_0+J_1]~,
\end{equation}
in light-cone coordinates
\begin{equation}
\partial _{\pm }J_{\mp }=\pm {\textstyle \frac 12}[J_{-},J_{+}]~,
\end{equation}
which leads to the following sequence of conservation laws for arbitrary
integers $m,n$:
\begin{equation}
\partial _{-}\hbox{Tr}~J_{+}^n=0=\partial _{+}\hbox{Tr}~J_{-}^m=0~.
\end{equation}
Such conservation laws are normally nontrivial ($n=m=2$ is energy-momentum
conservation), but for orthogonal groups such as exemplified here odd powers
vanish identically by virtue of the cyclicity of the trace.

\section{Poisson Brackets and Current Algebra}

Our goals in this section are to systematically work out the canonical
bracket algebra of all the local currents for the pseudodual theory, and,
through the use of these, to also demonstrate unequivocably that the action
of the charge for the nonlocal current $N_\mu$ given above is genuinely
nonlocal. Explicitly,
\begin{equation}
{}~ [\kern-0.45em\llap~[~ Q_N,\phi ^{ab}(y)~ ]\kern-0.45em\llap~]~ =-\bigl[%
[M^{ab},\phi(y)],\phi (y)\bigr]
+2\int_{-\infty}^{+\infty}dx~\varepsilon (y-x)[Z_0(x),M^{ab}]  \;
\end{equation}
will eventuate. Evidently then, $Q_N$ changes the boundary condition on the
field $\phi$ at $x=+\infty $ to a different one than at $x=-\infty $, and
thus changes the topology of the field configuration upon which it acts, a
change which is quantified by the charge $Q$.

First, observe that, for the PCM,
\begin{equation}
\label{J'sAsPhiAndPi}J_0(x)=\partial _x\phi (x),~~~~J_1(x)=\pi (x)-\frac
13[\partial _x\phi (x),\phi (x)]\;.
\end{equation}
So the $J$ current algebra follows immediately from the fundamental Poisson
brackets
\begin{equation}
{}~[\kern-0.45em\llap~[~\phi (x),\pi ^{ab}(y)~]\kern-0.45em\llap~%
]~=M^{ab}~\delta (x-y)=-~[\kern-0.45em\llap~[~\pi (x),\phi ^{ab}(y)~]\kern%
-0.45em\llap~]~,
\end{equation}
where we have suppressed one pair of indices, and introduced the O(N)
antisymmetrizer matrix
\begin{equation}
(M^{ab})_{cd}\equiv \delta _{ac}\delta _{bd}-\delta _{ad}\delta _{bc}.
\end{equation}
Note the distinction between matrix commutators, as given by $%
[~\cdots,\cdots~]$, and Poisson brackets, as given by $~[\kern-0.45em\llap~%
[~\cdots,\cdots~]\kern-0.45em\llap~]~$. Combining the fundamental brackets
with (\ref{J'sAsPhiAndPi}), we compute in succession
\begin{equation}
{}~[\kern-0.45em\llap~[~J_0(x),\phi ^{ab}(y)~]\kern-0.45em\llap~]~=0,~~~~~%
{}~[\kern-0.45em\llap~[~J_0(x),\pi ^{ab}(y)~]\kern-0.45em\llap~%
]~=M^{ab}~\delta ^{\prime }(x-y),~~~~~
{}~[\kern-0.45em\llap~[~J_0(x),[\partial _y\phi (y),\phi (y)]^{ab}~]\kern%
-0.45em\llap~]~=0,
\end{equation}
followed by
\begin{equation}
{}~[\kern-0.45em\llap~[~J_1(x),\phi ^{ab}(y)~]\kern-0.45em\llap~]~=-\delta
(x-y)~M^{ab}=-~[\kern-0.45em\llap~[~\phi (x),J_1^{ab}(y)~]\kern-0.45em\llap~%
]~,
\end{equation}
\begin{equation}
{}~[\kern-0.45em\llap~[~J_1(x),\pi ^{ab}(y)~]\kern-0.45em\llap~]~=-\frac
13[J_0(x),M^{ab}]~\delta (x-y)+\frac 13[\phi (x),M^{ab}]~\delta ^{\prime
}(x-y),
\end{equation}
\begin{equation}
{}~[\kern-0.45em\llap~[~J_1(x),[\partial _y\phi (y),\phi (y)]^{ab}~]\kern%
-0.45em\llap~]~=[J_0(y),M^{ab}]~\delta (x-y)+[\phi (y),M^{ab}]~\delta
^{\prime }(x-y)=~[\kern-0.45em\llap~[~J_1(x),[J_0(y),\phi (y)]^{ab}~]\kern%
-0.45em\llap~]~.
\end{equation}
Now, substitution of these results into the expressions in Eqn(\ref
{J'sAsPhiAndPi}) leads to the ``topological'' current algebra\footnote{%
N.B. Recall the distribution lemma $f(x,y)~ \delta^\prime(x-y)
=-\delta(x-y)~ \partial_x f(x,y)$ when $f(x,x)=0$. Also note, when $\phi$ is
an antisymmetric matrix, $[M^{ab},\phi] \equiv
\phi^{ac}M^{cb}-M^{ac}\phi^{cb}$, etc. Alternatively, $[M^{ab},\phi]_{cd} =
- [M^{cd},\phi]_{ab}$, and similarly, for antisymmetric $\phi$ and $\psi$, $%
[[M^{ab},\phi],\psi]_{cd} = [[M^{cd},\psi],\phi]_{ab}$.}
\begin{equation}
{}~[\kern-0.45em\llap~[~J_0(x),J_0^{ab}(y)~]\kern-0.45em\llap~]~=0,
\end{equation}
\begin{equation}
{}~[\kern-0.45em\llap~[~J_1(x),J_1^{ab}(y)~]\kern-0.45em\llap~%
]~=-[J_0(x),M^{ab}]~\delta (x-y),
\end{equation}
\begin{equation}
{}~[\kern-0.45em\llap~[~J_0(x),J_1^{ab}(y)~]\kern-0.45em\llap~]~=M^{ab}~\delta
^{\prime }(x-y)=~[\kern-0.45em\llap~[~J_1(x),J_0^{ab}(y)~]\kern-0.45em\llap~%
]~.
\end{equation}

Moving right along, bracket $Z_0$ with the local fields and topological
currents. First, act on the field and its conjugate. (Recall $%
Z_0=J_1+\frac12 [J_0,\phi]$.)
\begin{equation}
\label{Z0Phi}~ [\kern-0.45em\llap~[~ Z_0(x), \phi^{ab}(y)~ ]\kern-0.45em%
\llap~]~ = - M^{ab}~\delta(x-y) = - ~ [\kern-0.45em\llap~[~ \phi(x),
Z_0^{ab}(y)~ ]\kern-0.45em\llap~]~ , \;
\end{equation}
\begin{eqnarray}
\label{Z0Pi}~ [\kern-0.45em\llap~[~ Z_0(x), \pi^{ab}(y)~ ]\kern-0.45em\llap~%
]~ &=& \frac16 [J_0(x),M^{ab}]~\delta(x-y) - \frac16
[\phi(x),M^{ab}]~\delta^\prime(x-y) \nonumber
\\
&=&\frac13 \delta(x-y)~[J_0(x),M^{ab}] - \frac16 \partial_x \bigl( %
\delta(x-y)~[\phi(x),M^{ab}]\bigr) ~ ,
\end{eqnarray}
In the last step, we have isolated a total spatial derivative to make
transparent the action of the $Q_Z$ charge on $\pi$. We will often take such
clarifying steps below.

Next act on the topological current components.
\begin{equation}
\label{Z0J0}~ [\kern-0.45em\llap~[~ Z_0(x), J_0^{ab}(y)~ ]\kern-0.45em\llap~%
]~ = M^{ab} ~\delta^\prime(x-y) =~ [\kern-0.45em\llap~[~ J_0(x),
Z_0^{ab}(y)~ ]\kern-0.45em\llap~]~ , \;
\end{equation}
\begin{eqnarray}
\label{Z0J1}~ [\kern-0.45em\llap~[~ Z_0(x), J_1^{ab}(y)~ ]\kern-0.45em\llap~%
]~& =& - \frac12 [J_0,M^{ab}]~\delta(x-y) - \frac12
[\phi(x),M^{ab}]~\delta^\prime(x-y) \nonumber
\\
&=& - \frac12 \partial_x \bigl( \delta(x-y)~[\phi(x),M^{ab}]\bigr)
= - \frac12 [J_0,M^{ab}]~\delta(x-y) - ~ [\kern-0.45em\llap~[~ J_1(x),
Z_0^{ab}(y)~ ]\kern-0.45em\llap~]~ , \;
\end{eqnarray}
\begin{equation}
\label{J1Z0}~ [\kern-0.45em\llap~[~ J_1(x), Z_0^{ab}(y)~ ]\kern-0.45em\llap~%
]~ = - \frac12 [J_0,M^{ab}]~\delta(x-y) + \frac12
[\phi(y),M^{ab}]~\delta^\prime(x-y) = \frac12
[\phi(x),M^{ab}]~\delta^\prime(x-y) .
\end{equation}

Continuing this pattern, now bracket with $Z_1$. (Recall $Z_1=J_0+\frac12
[J_1,\phi]$.)
\begin{equation}
\label{Z1Phi}~ [\kern-0.45em\llap~[~ Z_1(x), \phi^{ab}(y)~ ]\kern-0.45em%
\llap~]~ = \frac12 \delta(x-y)~[\phi(x),M^{ab}] = ~ [\kern-0.45em\llap~[~
\phi(x), Z_1^{ab}(y)~ ]\kern-0.45em\llap~]~ , \;
\end{equation}
\begin{eqnarray}
\label{Z1Pi}~ [\kern-0.45em\llap~[~ Z_1(x), \pi^{ab}(y)~ ]\kern-0.45em\llap~%
]~& =& \Bigl( \frac12 [J_1(x),M^{ab}] +\frac16 [\phi(x), [J_0(x),M^{ab}]
]\Bigr) \delta(x-y) +\Bigl(M^{ab}-\frac16 [\phi(x),
[\phi(x),M^{ab}]]\Bigr)~\delta^\prime(x-y) \nonumber  \\
&=&\delta(x-y)~\Bigl( \frac12 [J_1(x),M^{ab}] +\frac13 [\phi(x),
[J_0(x),M^{ab}] ] +\frac16 [J_0(x),[\phi(x),M^{ab}] ] \Bigr) \nonumber \\
&+& \partial_x \Bigl( \delta(x-y)~\bigl(M^{ab} -\frac16 [\phi(x),
[\phi(x),M^{ab}]]\bigr) \Bigr) , \;
\end{eqnarray}
\begin{equation}
\label{Z1J0}~ [\kern-0.45em\llap~[~ Z_1(x), J_0^{ab}(y)~ ]\kern-0.45em\llap~%
]~ = - \frac12 [\phi(x),M^{ab}]~\delta^\prime(x-y) = \frac12
\delta(x-y)~[J_0(x),M^{ab}] - \frac12 \partial_x \bigl( \delta(x-y)~[%
\phi(x),M^{ab}]\bigr) , \;
\end{equation}

\begin{equation}
\label{Z1J1}~ [\kern-0.45em\llap~[~ Z_1(x), J_1^{ab}(y)~ ]\kern-0.45em\llap~%
]~ = \Bigl( \frac12 [J_1(x),M^{ab}] +\frac12 [\phi(x), [J_0(x),M^{ab}] ]
\Bigr)~\delta(x-y) + M^{ab}~\delta^\prime(x-y) . \;
\end{equation}

Combining these, we arrive at the pseudoabelian ``shift'' current algebra.
\begin{equation}
\label{Z0Z0}~ [\kern-0.45em\llap~[~ Z_0(x), Z_0^{ab}(y)~ ]\kern-0.45em\llap~%
]~ = \frac12 ~[J_0,M^{ab}] ~\delta(x-y), \;
\end{equation}
\begin{equation}
\label{Z1Z1}~ [\kern-0.45em\llap~[~ Z_1(x), Z_1^{ab}(y)~ ]\kern-0.45em\llap~%
]~ = \delta(x-y)~\Bigl( \frac12 [Z_1,M^{ab}] +\frac14
[\phi,[J_0,[\phi,M^{ab}]]] \Bigr), \;
\end{equation}
\begin{eqnarray}
\label{Z0Z1}~ [\kern-0.45em\llap~[~ Z_0(x), Z_1^{ab}(y)~ ]\kern-0.45em\llap~%
]~ &=& \Bigl( \frac12 [J_1,M^{ab}] - \frac14 [J_0,[\phi,M^{ab}]]
\Bigr)~\delta(x-y) +\Bigl( M^{ab} - \frac14 [\phi(x),[\phi(y),M^{ab}]]
\Bigr)~\delta^\prime(x-y) \nonumber \\
&=& \frac12 \delta(x-y)~[J_1(x),M^{ab}] + \partial_x \Bigl( \delta(x-y)~
\bigl(%
M^{ab}-\frac14 [\phi(x),[\phi(y),M^{ab}]] \bigr) \Bigr). \;
\end{eqnarray}
\begin{equation}
\label{Z1Z0}~ [\kern-0.45em\llap~[~ Z_1(x), Z_0^{ab}(y)~ ]\kern-0.45em\llap~%
]~ = \Bigl( \frac12 [J_1,M^{ab}] + \frac14 [\phi,[J_0,M^{ab}]]
\Bigr)~\delta(x-y) +\Bigl( M^{ab} - \frac14 [\phi(x),[\phi(y),M^{ab}]]
\Bigr)~\delta^\prime(x-y)
\end{equation}
$$
=\delta(x-y)~\Bigl( \frac12 [J_1(x),M^{ab}] +\frac14
[\phi,[J_0,M^{ab}]]+\frac14 [J_0,[\phi,M^{ab}]] \Bigr) + \partial_x \Bigl(
\delta(x-y)~ \bigl(M^{ab}-\frac14 [\phi(x),[\phi(y),M^{ab}]] \bigr) \Bigr).
\;
$$

Pressing onward, consider the remaining local current, $T_\mu$. (Recall $%
T_0=Z_1+[Z_0,\phi]$ and $T_1=Z_0+[Z_1,\phi]$.)
\begin{equation}
\label{T0Phi}~ [\kern-0.45em\llap~[~ T_0(x), \phi^{ab}(y)~ ]\kern-0.45em%
\llap~]~ = \frac32 [\phi,M^{ab}]~\delta(x-y) = ~ [\kern-0.45em\llap~[~
\phi(x), T_0^{ab}(y)~ ]\kern-0.45em\llap~]~ , \;
\end{equation}
\begin{equation}
\label{T0Pi}~ [\kern-0.45em\llap~[~ T_0(x), \pi^{ab}(y)~ ]\kern-0.45em\llap~%
]~ = \frac32 [\pi,M^{ab}]~\delta(x-y) + M^{ab}~\delta^\prime(x-y) = ~ [\kern%
-0.45em\llap~[~ \pi(x), T_0^{ab}(y)~ ]\kern-0.45em\llap~]~ , \;
\end{equation}
\begin{eqnarray}
\label{T0J0}~ [\kern-0.45em\llap~[~ T_0(x), J_0^{ab}(y)~ ]\kern-0.45em\llap~%
]~ &=& - \frac32 [\phi(x),M^{ab}]~\delta^\prime(x-y) = - ~ [\kern-0.45em\llap~%
[~ J_0(x), T_0^{ab}(y)~ ]\kern-0.45em\llap~]~ \nonumber \\
&=& \delta(x-y)~\frac32 [J_0(x),M^{ab}] -\partial_x \Bigl( \delta(x-y)~\frac32
[\phi(x),M^{ab}] \Bigr) , \;
\end{eqnarray}
\begin{eqnarray}
\label{T0J1}~ [\kern-0.45em\llap~[~ T_0(x), J_1^{ab}(y)~ ]\kern-0.45em\llap~%
]~ &=& \Bigl( [Z_0,M^{ab}]+\frac12 [J_1,M^{ab}]+[\phi,[J_0,M^{ab}]]
\Bigr)~\delta(x-y) +\Bigl( M^{ab}+\frac12 [\phi(x),[\phi(x),M^{ab}]]
\Bigr)~\delta^\prime(x-y) \nonumber \\
&=&\delta(x-y)~\frac32 [J_1,M^{ab}] +\partial_x \Bigl( \delta(x-y)~ \bigl(%
M^{ab}+\frac12 [\phi(x),[\phi(x),M^{ab}]]\bigr) \Bigr), \;
\end{eqnarray}
\begin{eqnarray}
\label{T0Z0}~ [\kern-0.45em\llap~[~ T_0(x), Z_0^{ab}(y)~ ]\kern-0.45em\llap~%
]~ &=& \Bigl( \frac32 [Z_0,M^{ab}]-\frac14 [J_0,[\phi,M^{ab}]]
\Bigr)~\delta(x-y) +\Bigl( M^{ab}-\frac14 [\phi(x),[\phi(y),M^{ab}]]
\Bigr)~\delta^\prime(x-y) \nonumber \\
&=&\delta(x-y)~\frac32 [Z_0,M^{ab}] +\partial_x \Bigl( \delta(x-y)~ \bigl(%
M^{ab}-\frac14 [\phi(x),[\phi(y),M^{ab}]]\bigr) \Bigr), \;
\end{eqnarray}
\begin{eqnarray}
\label{T0Z1}~ [\kern-0.45em\llap~[~ T_0(x), Z_1^{ab}(y)~ ]\kern-0.45em\llap~%
]~ &=& \Bigl( \frac32 [Z_1,M^{ab}]-[J_0,M^{ab}]+\frac14
[\phi,[J_0,[\phi,M^{ab}]]] +\frac14 [J_0,[\phi,[\phi,M^{ab}]]]
\Bigr)~\delta(x-y) \nonumber \\
&+&\Bigl(-[\phi(x),M^{ab}]+\frac14 [\phi(x),[\phi(x),[\phi(y),M^{ab}]]] \Bigr)
{}~\delta^\prime(x-y) \nonumber \\
&=&\delta(x-y)~\frac32 [Z_1,M^{ab}] +\partial_x \Bigl(\delta(x-y)~ \bigl(%
-[\phi(x),M^{ab}]+\frac14 [\phi(x),[\phi(x),[\phi(y),M^{ab}]]] \bigr)
\Bigr) , \;
\end{eqnarray}
\begin{equation}
\label{T0T0}~ [\kern-0.45em\llap~[~ T_0(x), T_0^{ab}(y)~ ]\kern-0.45em\llap~%
]~ = \frac32 [T_0(x),M^{ab}]~\delta(x-y)
\end{equation}
\begin{eqnarray}
\label{T0T1}~ [\kern-0.45em\llap~[~ T_0(x), T_1^{ab}(y)~ ]\kern-0.45em\llap~%
]~ &=& \Bigl( \frac32 [T_1,M^{ab}]-\frac54 [J_0,[\phi,M^{ab}]] +\frac14
[\phi,[J_0,[\phi,[\phi,M^{ab}]]]] +\frac14 [J_0,[\phi,[\phi,[\phi,M^{ab}]]]]
\Bigr)~\delta(x-y)  \nonumber \\
&+&\Bigl(M^{ab}-\frac54 [\phi(x),[\phi(y),M^{ab}]] +\frac14
[\phi(x),[\phi(x),[\phi(y),[\phi(y),M^{ab}]]]] \Bigr) ~\delta^\prime(x-y)
\end{eqnarray}      $$
=\delta(x-y)~\frac32 [T_1,M^{ab}] +\partial_x \Bigl(\delta(x-y)~\bigl(%
M^{ab}-\frac54 [\phi(x),[\phi(y),M^{ab}]] +\frac14
[\phi(x),[\phi(x),[\phi(y),[\phi(y),M^{ab}]]]] \bigr) \Bigr).
$$

Finally, consider the genuinely nonlocal current, $N_\mu$. (Recall $N_0 =
\frac12 [Z_1,\phi]+\frac12 [T_0,\phi]+[Z_0,\chi_Z]$ and $N_1 = \frac12
[Z_0,\phi]+\frac12 [T_1,\phi]+[Z_1,\chi_Z]$.) It suffices here to consider
only the time component.
\begin{equation}
\label{N0Phi}~ [\kern-0.45em\llap~[~ N_0(x), \phi^{ab}(y)~ ]\kern-0.45em%
\llap~]~ = \Bigl( [\chi_Z,M^{ab}]-[\phi, [\phi,M^{ab}]] \Bigr)~\delta(x-y)
-[Z_0(x),M^{ab}]~\theta(x-y) . \;
\end{equation}
This is sufficient to infer the action on $\phi$ of the nonlocal charge $Q_N
\equiv \int_{-\infty}^{+\infty}dx~N_0(x)$.
\begin{equation}
{}~ [\kern-0.45em\llap~[~ Q_N,\phi ^{ab}(y)~ ]\kern-0.45em\llap~]~ =-\bigl[%
[M^{ab},\phi(y)],\phi (y)\bigr]
+2\int_{-\infty}^{+\infty}dx~\varepsilon (y-x)[Z_0(x),M^{ab}], \;
\end{equation}
as claimed at the beginning of this section.

Actually, it is not too difficult to extend this result for the Poisson
bracket of the local field $\phi$ with the first genuine nonlocal charge to
the full set of nonlocal charges as contained in the path-ordered generating
functional $\chi$ of the previous section. To that end, we define the
path-ordered exponential for an interval $[x,y]$ (suppressing the $\kappa$%
-dependence which is understood to be carried by $C_1$)
\begin{equation}
\chi [x,y]=P\exp \Bigl(-\int_y^xdz~C_1(z,t)\Bigr),
\end{equation}
and note the general relation for the variation of $\chi$ induced by any
variation of the matrix $C_1$, such as that obtained from a Poisson bracket,
\begin{equation}
\delta\chi [x,y]=-\int_y^xdz~\chi [x,z]~\delta C_1(z)\chi [z,y].
\end{equation}
To determine the full nonlocal transformations of $\phi $ and $\pi $ we
therefore utilize their Poisson brackets with $\chi $, or in view of the
last relation, with $C_1$. These are given by
\begin{equation}
{}~[\kern-0.45em\llap~[~C_0(x),\phi ^{ab}(y)~]\kern-0.45em\llap~]~={\frac
\kappa {1-\kappa ^2}}~\delta (x-y)~M^{ab},
\end{equation}
\begin{equation}
{}~[\kern-0.45em\llap~[~C_1(x),\phi ^{ab}(y)~]\kern-0.45em\llap~]~={\frac{%
\kappa ^2}{1-\kappa ^2}}~\delta (x-y)~M^{ab},
\end{equation}
\begin{equation}
{}~[\kern-0.45em\llap~[~C_0(x),\pi ^{ab}(y)~]\kern-0.45em\llap~]~={\frac{%
-\kappa }{1-\kappa ^2}}~\Bigl(-{\frac 23}\delta
(x-y)~[J_0(x),M^{ab}]+\partial _x\bigl( \delta (x-y)~(\kappa M^{ab}+{\ \frac
13}[\phi (x),M^{ab}])\bigr)
\Bigr),
\end{equation}
\begin{equation}
{}~[\kern-0.45em\llap~[~C_1(x),\pi ^{ab}(y)~]\kern-0.45em\llap~]~={\frac{%
-\kappa }{1-\kappa ^2}}~\Bigl(-{\frac 23}\kappa \delta
(x-y)~[J_0(x),M^{ab}]+\partial _x\bigl( \delta (x-y)~(M^{ab}+{\ \frac 13}%
\kappa [\phi (x),M^{ab}])\bigr)
\Bigr).
\end{equation}

Thus, the full nonlocal transformations of the field and its conjugate
momentum are given by
\begin{equation}
\delta _\kappa \phi ^{ab}(x)\equiv {\frac 12}~[\kern-0.45em\llap~[~\hbox{Tr}~%
\bigl(\Omega \chi [\infty ,-\infty ]\bigr),\phi ^{ab}(x)~]\kern-0.45em\llap~%
]~={\frac{-{\frac 12}\kappa ^2}{1-\kappa ^2}} \hbox{Tr}~\bigl(\Omega \chi
[\infty ,x]~M^{ab}~\chi [x,-\infty ]\bigr),
\end{equation}
\begin{equation}
\delta _\kappa \pi ^{ab}(x)={\frac{-{\frac 12}\kappa ^2}{1-\kappa ^2}}
\hbox{Tr}~\Bigl(\Omega \chi [\infty ,x]~\bigl(
[{\frac 23}J_0(x)+{\frac 1\kappa }C_1(x),M^{ab}]-{\frac 13}%
[C_1(x),[M^{ab},\phi (x)]]\bigr)
{}~\chi [x,-\infty ]\Bigr).
\end{equation}
Also note that
\begin{equation}
{}~[\kern-0.45em\llap~[~C_0(x),[\partial _y\phi (y),\phi (y)]^{ab}~]\kern%
-0.45em\llap~]~={\frac{-\kappa }{1-\kappa ^2}}~\Bigl(\delta
(x-y)~[J_0(x),M^{ab}]+\partial _x\bigl( \delta (x-y)~[\phi (x),M^{ab}])%
\bigr)
\Bigr),
\end{equation}
\begin{equation}
{}~[\kern-0.45em\llap~[~C_1(x),[\partial _y\phi (y),\phi (y)]^{ab}~]\kern%
-0.45em\llap~]~=\kappa ~~[\kern-0.45em\llap~[~C_0(x),[\partial _y\phi
(y),\phi (y)]^{ab}~]\kern-0.45em\llap~]~,
\end{equation}
which, when substituted into $\delta \chi$, yields the nonlocal
transformation of $[\partial _x\phi ,\phi ]$ .

We may then combine all these results to obtain the full nonlocal
transformation of the local currents. We find the simple result (note that $%
C_1+\kappa J_0=\kappa C_0$)
\begin{equation}
\label{deltaJ}\delta _\kappa J_\mu ^{ab}(x)={\frac 12}~[\kern-0.45em\llap~[~%
\hbox{Tr}~\bigl(\Omega \chi [\infty ,-\infty ]\bigr),J_\mu ^{ab}(x)~]\kern%
-0.45em\llap~]~={\frac{-{\frac 12}\kappa ^2}{1-\kappa ^2}} \hbox{Tr}~\bigl(
\Omega \chi [\infty ,x]~[\tilde C_\mu (x),M^{ab}]~\chi [x,-\infty ]\bigr),
\;
\end{equation}
where all fields are evaluated at the same time.

\section{Concluding Remarks}

The canonical machinery of the last section permits us to address one final
point in conclusion. Davies et al. \cite{MacFarlaneandCo} have criticized
the use of Noetherian methods to generate the nonlocal currents through
nonlocal variations of fields, such as used in \cite{CurtrightZachos'81} for
the Gross-Neveu model. They argue that the proper demonstration that the
nonlocals constitute meaningful conservation laws, even for the classical
theory, must consist in showing through canonical methods that they are
constants of the motion. That is, for the classical theory, the nonlocal
charges must have vanishing Poisson brackets with the Hamiltonian, while for
the quantum theory, they must have vanishing commutators.

At least for the classical theory, the result (\ref{deltaJ}) leads to this
result, and more. It immediately shows that the Hamiltonian {\em density} is
invariant under the full nonlocal symmetry.
\begin{equation}
\delta _\kappa {\cal H}=0,
\end{equation}
where
\begin{equation}
{\cal H}={\textstyle \frac 12} ~\hbox{Tr}~\Bigl(J_0J_0+J_1J_1\Bigr),
\end{equation}
since
\begin{equation}
[J_0,C_1]=-[J_1,C_0]={\frac{-\kappa ^2}{1-\kappa ^2}}[J_0,J_1].
\end{equation}
Furthermore, (\ref{deltaJ}) also shows that the momentum density is
invariant under the nonlocal transformations.
\begin{equation}
\delta _\kappa {\cal P}=0,
\end{equation}
where
\begin{equation}
{\cal P}=\hbox{Tr}~\Bigl(J_0J_1\Bigr),
\end{equation}
since
\begin{equation}
[J_0,C_0]=-[J_1,C_1]={\frac{-\kappa }{1-\kappa ^2}}[J_0,J_1].
\end{equation}
Moreover, because the energy-momentum tensor is symmetric and traceless for
the classical theory, these two results lead to the conclusion that all
components of the classical local energy-momentum tensor are invariant under
the nonlocal symmetry.
\begin{equation}
\delta _\kappa \theta _{\mu \nu }=0,
\end{equation}
\begin{equation}
\theta _{\mu \nu }=\hbox{Tr}\Bigl(J_\mu J_\nu -{\textstyle \frac 12}
g_{\mu \nu }J_\lambda J^\lambda \Bigr).
\end{equation}
Thus we have established within the canonical framework the conservation of
all the classical nonlocal charges as contained in the master charge
generating functional.

A next logical step would be to consider the status of this classical result
in the context of the quantum theory. This step will be taken in a
subsequent investigation. Suffice it to say here that quantum corrections
are indeed expected in the transformation of the local energy-momentum
tensor. In particular, since the trace of that tensor, $\theta _\mu ^\mu$,
probably does not vanish in the quantum theory, as suggested by the
nonvanishing one-loop renormalization flow, the nonlocal symmetry will
probably not leave the untraced local energy-momentum tensor invariant in
the quantum theory \cite{LeClairandCo}. However, since the PCM is {\em not}
asymptotically free, a reliable short-distance expansion is not within the
grasp of direct perturbative methods for this model. Therefore, a convincing
analysis of nonlocal transformations for the pseudodual quantum theory may
take quite some time.

\acknowledgements
Work supported by the NSF grant
PHY-92-09978 and
the U.S.~Department of Energy, Division of High Energy Physics, Contract
W-31-109-ENG-38. T. Curtright thanks FNAL for its hospitality during
a significant
portion of this project.

\end{document}